\documentclass[aps,pra,twocolumn,superscriptaddress,showpacs]{revtex4-2}
\usepackage{indentfirst}
\usepackage{graphicx}
\usepackage{dcolumn}
\usepackage{bm}
\usepackage{amsmath}
\usepackage{amssymb}
\usepackage{latexsym}
\usepackage{epsfig}
\usepackage{amsbsy}
\usepackage{array}
\usepackage{amssymb}
\usepackage{setspace}
\usepackage{bm}
\usepackage{endnotes}
\usepackage{indentfirst}
\begin{document}
\preprint{Physical Review A}
\title{Shortened Effect of Coherence Length of Light due to Nonselective Linear Absorption}
\author{Xingchu Zhang}
\thanks{These two authors contributed equally to this work.}
\affiliation{School of Physics and Information Engineering, Guangdong University of Education, Guangzhou 510303, China }
\author{Zhencheng Huang}
\thanks{These two authors contributed equally to this work.}
\author{Zidong Liang}
\author{Weilong She }
\email[]{shewl @mail.sysu.edu.cn}

\affiliation{School of Physics, Sun Yat-Sen University, Guangzhou 510275, China.}

\begin{abstract}
From the Michelson interference of He-Ne laser beam, it is found that the coherence length of the beam decreases with the decrease of intensity when the laser beam passes through a nonselective absorption filter and the intensity becomes low enough. The experiment verifies the characteristic of discrete wavelet structure of light for the first time.
\end{abstract}


\maketitle
\section{INTRODUCTION}
The coherence\cite{1,2,3} is an important property of light, which is the foundation of many applications such as holography\cite{4}, phase contrast microscopy\cite{5}, optical coherence tomography\cite{6} and gravitational wave detection\cite{7}. In most situations, the coherence of light is determined by the radiation source and depends on the process of light generation\cite{8,9}. Here, we pay our attention to the temporal coherence of light. The coherence length is generally used to describe the temporal coherence of  a light wave and is closely related to the monochromaticity of light\cite{1}. As is well known, the light from a source of spontaneous radiation has a short coherence length and poor monochromaticity, while a laser beam has a long coherence length and is with good monochromaticity. In the cavity, there are various methods for improving the monochromaticity of laser spectral line\cite{10,11,12,13,14,15}. And outside the source, besides bandpass filter that can improve the monochromaticity of light, the nonlinear optical effect, such as self-phase-modulated \cite{16,17}, is found to be able to broaden the spectrum of laser beam that means shortening the coherence length. However, as far as we know, there is no experimental report about the effect of nonselective linear absorption on the coherence length of laser beam outside the cavity. Recently, Zhang and She developed a discrete wavelet structure theory of classic plane light waves\cite{18}. According to their theory, a classic plane light wave of finite length can be decomposed into a series of discrete wavelets and is with discrete energy. If it is the case, one would expect that, for a light wave with finite coherence length, the nonselective linear absorption would lead to the reduction of the coherence length when the light intensity becomes low enough. By the Michelson interference of He-Ne laser beam, we found that the coherence length of the beam decreases with the decrease of intensity when the laser passes through a nonselective absorption filter and the intensity becomes low enough. This is consistent with our expectation. 
   
\section{EXPERIMENTAL RESULTS}
\begin{figure*}
\includegraphics[height=9.0in,width=5.52in]{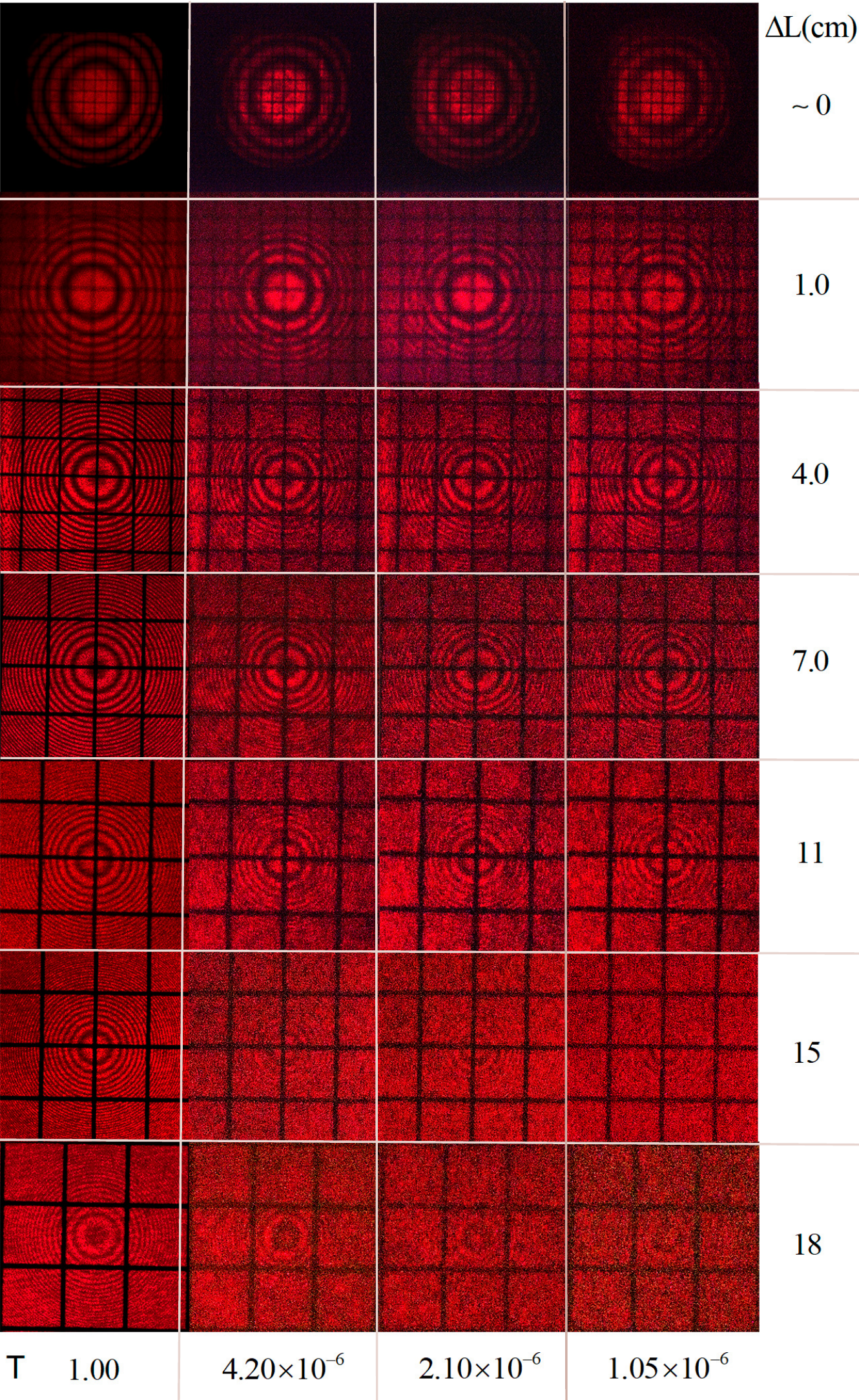}
\caption{\label{fig:wide} Michelson interference experiment on the shortened effect of coherence length of light due to linear absorption, where $\Delta L$ is the optical path difference between the two arms of the interferometer, corresponding to the row; $T$ is the transmittance of the attenuator, corresponding to column, and $T=1$ is the special case without attenuator.}
\end{figure*}

\begin{figure*}
\includegraphics[height=9.0in,width=5.52in]{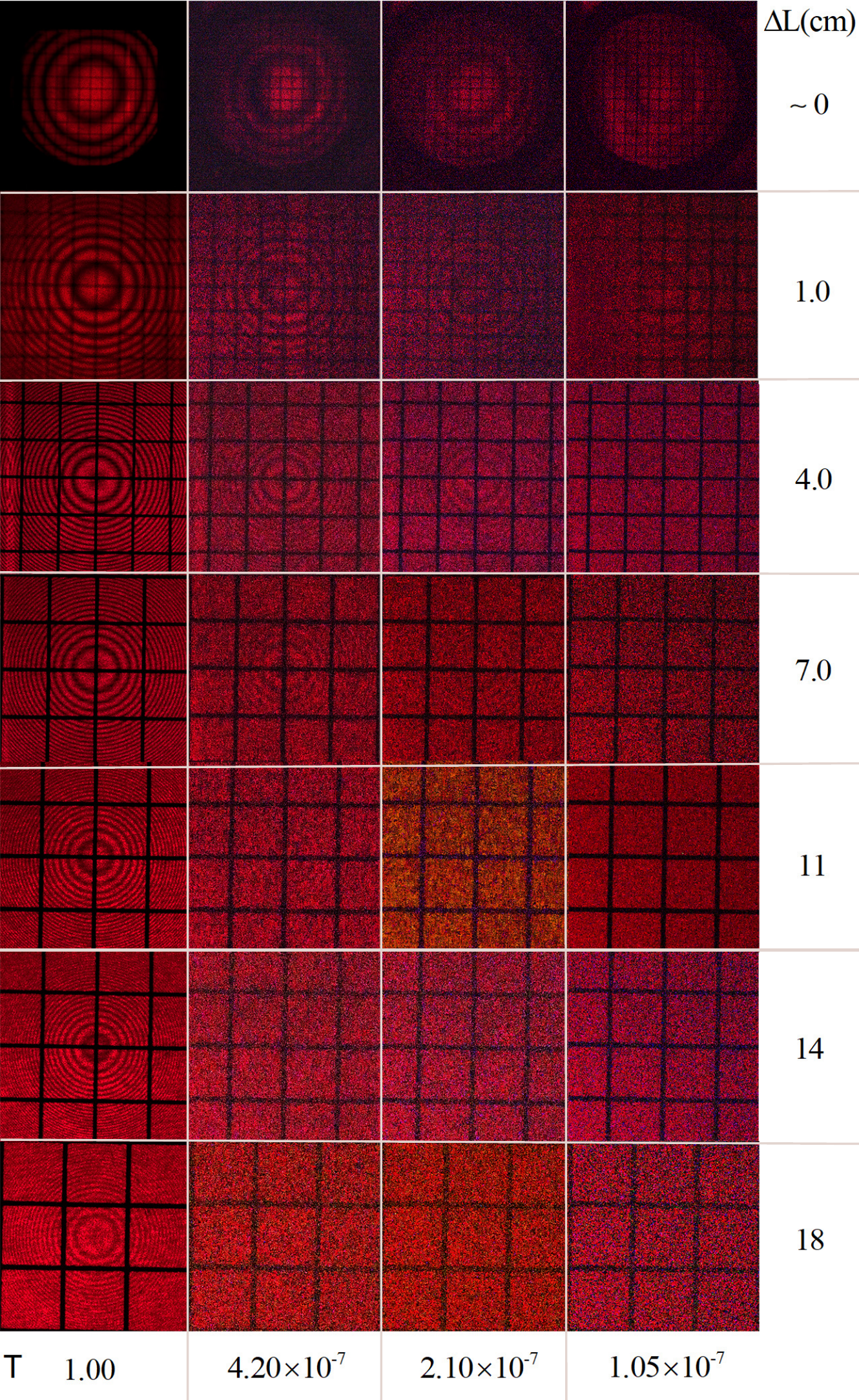}
\caption{\label{fig:wide}  Michelson interference experiment on the shortened effect of coherence length of light due to linear absorption, where $\Delta L$ is the optical path difference between the two arms of the interferometer, corresponding to the row; $T$ is the transmittance of the attenuator, corresponding to column, and $T=1$ is the special case without attenuator.}
\end{figure*}

Our experiment is carried out with a Michelson interferometer. The light source is a He-Ne laser with an expander placed near the laser exit port. The laser power behind the expander is 2.53 $mW$, denoted by $P_{0}$, the coherence length of laser beam is measured to be about 22 $cm$, denoted by $L_{0}$. The laser beam is split by a prism beam splitter (BS). An attenuator, made of nonselective absorption filter, is placed between the expander and BS, but much near to the expander. The diameter of the laser beam at the filter is about 1.7 $cm$. So, the initial intensity $I_{0}$ for the filter is about $1.115 \times {10^{ - 3}}$ $W/c{m^2}$.  Two laser beams from BS are reflected by mirrors M1 and M2, respectively, then one of them passes through the prism and another one is reflected by the prism again. The two beams from prism therefore go overlapping, then interfere with each other. The interference fringes are projected on a screen with grids for observation. The use of grids is for convenience of focusing. During the whole experiment, the interference fringes are adjusted as an equal inclination pattern and taken by a camera (model PENTAX K-3). The exposure time of photography is fixed at 1/200 $s$  when without attenuator. And when with attenuator, unless otherwise stated specially, the product of the light power after the attenuator and the exposure time is set to be the same value. Denote transmittance of the attenuator as $T$ and the exposure time as $\Delta t$, respectively, then $P_{0}{\Delta t}T=constant$. After many trials, this constant is chosen as $1.594 \times {10^{ - 7}} J$. We have performed a lot of experiments and found that when there is no significant attenuation of the light intensity, for example, when $T > 4.20 \times {10^{ - 6}}$, the change of the coherence length is so small that it cannot be observed. However, when the laser beam passes through the attenuator and the intensity becomes low enough, the coherence length will be shortened evidently and further decrease with the decrease of the intensity.  Here we give two sets of typical experiments. The first one is shown in Fig.1, where $\Delta L$ = $ 0,1,4,7,11,15,18 {\kern 1pt} {\kern 1pt} {\kern 1pt} cm$ are the optical path differences between the two arms of the interferometer, corresponding respectively to the rows in the figure; and  $T = 1.00,4.20 \times {10^{ - 6}},2.10 \times {10^{ - 6}},1.05 \times {10^{ - 6}}$ are the transmittances of the attenuator, corresponding respectively to the columns in the figure. Let's briefly explain columns 1-4. Column 1 is the case without attenuator that is for reference purpose, the transmittance for column 2 is $T = 4.20 \times {10^{ - 6}}$, and from columns 3 to 4 each transmittance decreases successively by half relative to the previous one. So, the exposure times are 15, 30, 60 $s$ for columns 2 to 4 (rows 1 to 6), respectively. And from column 2 to column 4 in row 7, the exposure times are specially prolongated, they are 20, 40 and 80 $s$, respectively. The second set of experiments is shown in Fig.2. The optical path differences to the rows, are the same as those in the first set. Besides column 1 without attenuator, the change way of transmittance from columns 2 to 4 is like that in the first set. The exposure times are: 150, 300, 600 $s$ for columns 2 to 4 (rows 1 to 4); 150, 300 $s$ for columns 2 to 3 (rows 5 and 6), respectively; while specially, 1000 $s$ for column 4 (row 5), 800 $s$ for column 4 (row 6), and 200, 400, 800 $s$ for columns 2 to 4 (row 7), respectively. The purpose of prolongating the exposure time in special cases is to prove that, with the increase of $\Delta L$ or the decrease of $T$, the becoming blurred of fringes is not due to underexposure. It should be noted that the backgrounds of the figures have been deducted. Let's look at the interference fringes in Fig.1. When $\Delta L < 15$  $cm$, although the attenuator takes different transmittances, the clearness of the interference fringes from columns 2 to 4 is almost the same. But when $\Delta L = 15$ $cm$, the interference fringe for $T = 1.05 \times {10^{ - 6}}$ (column 4) is slightly blurrier than that for $T = 2.10 \times {10^{ - 6}}$ (column 3). From the experiment, we find that as the optical path difference $\Delta L$ further increases, the difference of the fringes among different transmittances becomes obvious. For example, for $\Delta L=18$  $cm$, the visibility of interference fringes decreases with the decrease of the transmittance. And when  $T = 1.05 \times {10^{ - 6}}$, the interference fringes become quite blurred. It means that due to linear absorption the coherent length of light beam decreases with the decrease of the intensity. This can be seen more clearly in Fig.2, for which we can compare the interference fringes from two directions: row and column. Let's first see the column. For each $T$ value ($ \ne 1$), as $\Delta L$ increases (from row 1 to row 7), the interference fringes become blurred gradually and eventually disappear. It shows that due to linear absorption, the coherence length will be shortened and no longer maintain the original value of $22$ $cm$. Then see the row. For each $\Delta L$, as $T$ decreases (from column 2 to column 4), the interference fringes also become blurred gradually, showing that due to linear absorption, the coherence length of the beam decreases with the decrease of intensity. Contrast Fig.1 with Fig.2, for example, at $\Delta L=11$ $cm$, one can see this phenomenon very clearly. From Fig.1 and Fig.2, one can find that the coherence lengths corresponding to transmittances $1.05 \times {10^{ - 6}}$,$4.20 \times {10^{ - 7}}$,$2.10 \times {10^{ - 7}}$,$1.05 \times {10^{ - 7}}$ are about 18, 14, 11 and 7 $cm$, respectively. The changes of coherence length for $T = 4.20 \times {10^{ - 6}}$, $2.10 \times {10^{ - 6}}$ are too small to observe, so they are not obviously shown in Fig.1. The experiment will be explained below.  

\section{DISCUSSION}
The temporal coherence of light waves is usually explained in the time domain or frequency domain\cite{1}. In time domain, the cut cosine wave train model is often used for interpretation \cite{1}. But a section of cut cosine wave train is not the solution of Maxwell's equations. Here, we try to make an explanation on the above experiments by using the discrete wavelet structure theory of classic plane light waves proposed by Zhang and She\cite{18}. According to their theory, a plane light wave of finite length can be described by the following wave train with discrete wavelet structure:
\begin{eqnarray}
{E_k}(z - ct) = {E_{k0}}{e^{ik(z - ct)}}\sum\limits_{r =  - m}^{m + 1} {{e^{\frac{{ - {{[(z - ct) - r\lambda  + \lambda /2]}^2}}}{{2s}}}}},
\label{eq:energy}
\end{eqnarray}
where $E_{k0}$ is the amplitude of the light wave, $c$ is the speed of light in vacuum. $k = \omega /c$  is the wave vector ($\omega$  is the idler frequency) and $\lambda  = 2\pi /k$; $s = {(\lambda /{c_1})^2}/2$ with $c_{1}=0.886231921$. $m$ is the discrete wavelet structure parameter, which is an integer. Referring to the linear absorption law of light, we regard the laser beam as the superposition of many independent basic wave trains with amplitude $E_{k0\min}$ \cite{18}. Let $j$ be the number of basic wave trains, according to Ref.\cite{18}, the energy of the wave trains corresponding to the cross-section $S={\lambda ^2}$ can be expressed as $\varepsilon=mj{p_{0k}}\omega $, where $p_{0k}$ is a constant. Through the medium of cross-sectional area $S$, when the basic wave train is absorbed and loses a portion of energy ${p_{0k}}\omega $, the wavelet structure parameter $m$ will decrease by 1, and the length of the basic wave train will decrease by $2\lambda $ correspondingly. On the other hand, according to Bouguer-Lambert law $dI=-\alpha Idl{\kern 1pt} $, considering $d\varepsilon  = S\Delta \tau dI =  - S\Delta \tau \alpha Idl$ (${\Delta \tau}$  is a small time-duration). On account of $\varepsilon  = SI\Delta \tau $, so $d\varepsilon  =  - \alpha \varepsilon dl$. We can obtain $\varepsilon  = {\varepsilon _0}{e^{ - \alpha l}}$, then $mj = {m_0}{j_0}{e^{ - \alpha l}}$, and   
\begin{eqnarray}
\Delta (mj) =  - \alpha {m_0}{j_0}{e^{ - \alpha l}}\Delta l =  - \alpha mj\Delta l,
\label{eq:energy}
\end{eqnarray}
where $\Delta (mj) = j\Delta m + m\Delta j$. $j\Delta m$ corresponds to the energy change caused by the change of the wavelet structure parameters, and $m\Delta j$ corresponds to the energy change caused by the change of the number of basic wave trains. We try making a translation from  $m\Delta j$ into  $j\Delta m$. Let $m\Delta j = Aj\Delta m$, where $A$ is a function of energy. Let it be proportional to energy, i.e., $A \propto mj$, and we can write $m\Delta j = Bjmj\Delta m$, where $B$ is a scale coefficient. Then $\Delta (mj) = j\Delta m(1 + Bmj)$, and Eq.(2) can be rewritten as
\begin{eqnarray}
- \alpha mj\Delta l =j(1 + Bmj) \Delta m.
\label{eq:energy}
\end{eqnarray} 
That is $(1 + Bmj)\Delta m =  - \alpha m\Delta l$, then we can obtain 
\begin{eqnarray}
\Delta m = \frac{{ - \alpha m\Delta l}}{{1 + Bmj}} = \frac{{ - \alpha m\Delta l}}{{1 + B{m_0}{j_0}{e^{ - \alpha l}}}} < 0.
\label{eq:energy}
\end{eqnarray} 
It shows that due to absorption, the basic wave trains are shortened. For ${m_0}{j_0}$ represents the initial energy and is related to the initial light intensity $I_{0}$, so Eq. (4) can be changed to 
\begin{eqnarray}
\Delta m = \frac{{ - \alpha m\Delta l}}{{1 + C{I_0}{e^{ - \alpha l}}}},
\label{eq:energy}
\end{eqnarray} 
Or
\begin{eqnarray}
\frac{{\Delta m}}{m} = \frac{{ - \alpha \Delta l}}{{1 + C{I_0}{e^{ - \alpha l}}}},
\label{eq:energy}
\end{eqnarray}
where $C$ is a coefficient to be determined. From Eq.(6), one can easily get 
\begin{eqnarray}
m = \frac{{(1 + C{I_0}){e^{ - \alpha l}}}}{{1 + C{I_0}{e^{ - \alpha l}}}}{m_0} = \frac{{(1 + C{I_0})T}}{{1 + C{I_0}T}}{m_0}.
\label{eq:energy}
\end{eqnarray}
Assume that the initial coherent length of the laser beam is $L_{0}$, according to Ref.\cite{18}, at the front of the attenuator, the initial wavelet structure parameter of the beam is ${m_0} = {L_0}/2\lambda $. After the attenuator, the wavelet structure parameter of the light wave is changed to $m$, and the coherence length is correspondingly reduced to $L = m \cdot 2\lambda $. Therefore, we have
\begin{eqnarray}
L = \frac{{(1 + C{I_0})T}}{{1 + C{I_0}T}}{L_0}.
\label{eq:energy}
\end{eqnarray}
In Fig.2, from column 4 of row 4, where $T = 1.05 \times {10^{ - 7}}$, we find that the coherence length is about 7 $cm$. Then we can determine the constant $C$, which is $3.995 \times {10^9}$ ${W^{ - 1}}{{cm}^{2}}$. Using the transmittances $T = 4.20 \times {10^{ - 6}}$, $2.10 \times {10^{ - 6}}$, $1.05 \times {10^{ - 6}}$, $4.20 \times {10^{ - 7}}$, $2.10 \times {10^{ - 7}}$ and Eq.(8), we can further calculate the correspondent coherent lengths after the attenuator, which are $L = 20.9$, 19.9, 18.1, 14.3, 10.6 $cm$, respectively. This is consistent with the above observation. Note that the first two values of $T$ yield very little effect on the coherence length, although they make remarkable reduction of the  light intensity. In fact, in this situation, it is difficult to find the difference between $L$ and $L_{0}$ in experiment. This is the reason why the effect of nonselective linear absorption on the coherence length of light waves has never been observed under normal conditions. From the above discussion, one would find that if $I_{0}$ is low enough and  $T > 1$  in Eq.(8), i.e., the medium is with gain, the coherence length can be extended due to light amplification but it will reach saturation when $C{I_0}T >> 1$.

\section{CONCLUSION}
In conclusion, from Michelson interference of He-Ne laser beam, we have found that the nonselective linear absorption will reduce the coherence length of the beam, and the effect can be observed when the beam passes through a nonselective absorption filter and the intensity becomes low enough. The experiment verifies the characteristic of discrete wavelet structure of light for the first time.

\end{document}